\documentclass[aps, prl, twocolumn,superscriptaddress,preprintnumbers,amsmath,amssymb]{revtex4}
\usepackage[paperwidth=210mm,paperheight=297mm,centering,hmargin=2.1cm,vmargin=2.3cm]{geometry}
\usepackage{graphicx}
\usepackage{dcolumn}
\usepackage{bm}
\usepackage{siunitx}
\usepackage{float}
\usepackage{graphicx,here}

\usepackage{color}
\usepackage{soul}

\begin{document}

\title{Element-specific field-induced spin reorientation and \\an unusual tetracritical point in MnCr$_2$S$_4$}

\author{Sh. Yamamoto}
\email{s.yamamoto@hzdr.de}
\affiliation{Hochfeld-Magnetlabor Dresden (HLD-EMFL) and W\"{u}rzburg-Dresden Cluster of Excellence ct.qmat, Helmholtz-Zentrum Dresden-Rossendorf, 01328 Dresden, Germany}
\author{H. Suwa}
\affiliation{Department of Physics, The University of Tokyo, Tokyo 113-0033, Japan}
\author{T. Kihara}
\affiliation{Institute for Materials Research, Tohoku University, Sendai 980-8577, Japan}
\author{T. Nomura}
\affiliation{Hochfeld-Magnetlabor Dresden (HLD-EMFL) and W\"{u}rzburg-Dresden Cluster of Excellence ct.qmat, Helmholtz-Zentrum Dresden-Rossendorf, 01328 Dresden, Germany}
\affiliation{Institute for Solid State Physics, The University of Tokyo, Kashiwa, Chiba 277-8581, Japan}
\author{Y. Kotani}
\affiliation{Japan Synchrotron Radiation Research Institute, SPring-8, Sayo, Hyogo 679-5198, Japan}
\author{T. Nakamura}
\affiliation{Institute for Materials Research, Tohoku University, Sendai 980-8577, Japan}
\affiliation{Japan Synchrotron Radiation Research Institute, SPring-8, Sayo, Hyogo 679-5198, Japan}
\author{Y. Skourski}
\affiliation{Hochfeld-Magnetlabor Dresden (HLD-EMFL) and W\"{u}rzburg-Dresden Cluster of Excellence ct.qmat, Helmholtz-Zentrum Dresden-Rossendorf, 01328 Dresden, Germany}
\author{S. Zherlitsyn}
\affiliation{Hochfeld-Magnetlabor Dresden (HLD-EMFL) and W\"{u}rzburg-Dresden Cluster of Excellence ct.qmat, Helmholtz-Zentrum Dresden-Rossendorf, 01328 Dresden, Germany}
\author{L. Prodan}
\affiliation{Institute of Applied Physics, MD 2028, Chisinau, R. Moldova}
\affiliation{Experimental Physics 5, Center for Electronic Correlations and Magnetism,
Institute of Physics, University of Augsburg, 86159, Augsburg, Germany}
\author{V. Tsurkan}
\affiliation{Institute of Applied Physics, MD 2028, Chisinau, R. Moldova}
\affiliation{Experimental Physics 5, Center for Electronic Correlations and Magnetism,
Institute of Physics, University of Augsburg, 86159, Augsburg, Germany}
\author{H. Nojiri}
\affiliation{Institute for Materials Research, Tohoku University, Sendai 980-8577, Japan}
\author{A. Loidl}
\affiliation{Experimental Physics 5, Center for Electronic Correlations and Magnetism,
Institute of Physics, University of Augsburg, 86159, Augsburg, Germany}
\author{J. Wosnitza}
\affiliation{Hochfeld-Magnetlabor Dresden (HLD-EMFL) and W\"{u}rzburg-Dresden Cluster of Excellence ct.qmat, Helmholtz-Zentrum Dresden-Rossendorf, 01328 Dresden, Germany}
\affiliation{Institut f$\ddot{u}$r Festk$\ddot{o}$rper- und Materialphysik, TU Dresden, 01062 Dresden, Germany}

\date{\today}

\begin{abstract}

The ferrimagnetic spinel MnCr$_2$S$_4$ shows a variety of magnetic-field-induced phase transitions owing to bond frustration and strong spin-lattice coupling. However, the site-resolved magnetic properties at the respective field-induced phases in high magnetic fields remain elusive. Our soft x-ray magnetic circular dichroism studies up to 40 T directly evidence element-selective magnetic-moment reorientations in the field-induced phases. The complex magnetic structures are further supported by entropy changes extracted from magnetocaloric-effect measurements. Moreover, thermodynamic experiments reveal an unusual tetracritical point in the $H$-$T$ phase diagram of MnCr$_2$S$_4$ due to strong spin-lattice coupling.

\end{abstract}

\maketitle

Frustrated magnets with competing magnetic interactions offer an exceptional playground for studying a variety of exotic magnetic states and emergent magnetic excitations \cite{Lacroix2011introduction}. The competing interactions between the constituent magnetic ions can be fine-tuned by applying an external magnetic field, which stabilizes various field-induced states with complex magnetic structures \cite{willenberg2012magnetic,zvyagin2019pressure}. This leads to a rich magnetic-field-temperature ($H$-$T$) phase diagram that also can be used as a test bed for validating various theoretical models \cite{diep2013frustrated,kawamura1998universality}. Cubic spinels $AB_2X_4$, where $A$ and $B$ represent magnetic cations, belong to paradigmatic frustrated systems. Geometrical frustration results from the $B$ magnetic cation being located at the vertices of a pyrochlore lattice with antiferromagnetic (AFM) exchange interactions, whereas bond frustration is caused by the competing exchange interactions of $J_{A\text{-}A}$, $J_{A\text{-}B}$, and $J_{B\text{-}B}$ \cite{rudolf2007spin}. In addition to the magnetic frustration, a strong spin-lattice coupling gives rise to a robust magnetization plateau, to unconventional metastable magnetostructural \cite{tsurkan2013unconventional}, and to spin-driven multiferroic states \cite{wosnitza2016frustrated}. 

MnCr$_2$S$_4$ crystallizes in a normal cubic-type spinel structure (space group $Fd\overline{3}m$) where the Mn$^{2+}$ ($S$ = 5/2, 3$d^5$) and Cr$^{3+}$ ($S$ = 3/2, 3$d^3$) ions reside on the tetrahedral $A$ and octahedral $B$ sites, respectively. This results in a bond frustration with competing AFM interactions between $J_{\mathrm{Mn}\text{-}\mathrm{Cr}}$ and $J_{\mathrm{Mn}\text{-}\mathrm{Mn}}$. The title compound shows two consecutive transitions at $T_C$ $\approx$ 65 K to a ferrimagnetic state and at $T_\mathrm{YK}$ $\approx$ 5 K to a canted spin configuration \cite{tsurkan2003magnetic}. It is suggested that at $T_\mathrm{YK}$, the Yafet-Kittel (YK) phase is stabilized. This phase is composed of two nonequivalent manganese sites, Mn1 and Mn2, with the magnetic moments rotated by +120$^\circ$ and -120$^\circ$ with respect to the chromium moment that is parallel to the external magnetic field caused by the strong $J_{\mathrm{Cr}\text{-}\mathrm{Cr}}$ FM interaction \mbox{\cite{baltzer1966exchange}}. A variety of magneto-structural phases have been identified in the $H$-$T$ phase diagram spanning up to 100 T and below 20 K, including a robust plateau between 25-50 T \cite{miyata2020spin} (see Sec. I of the Supplemental Material \mbox{\cite{yamamoto2020SM}}). In addition, from a quantum lattice-gas model, the YK state and the intermediate phase, which is located between the YK and plateau phases, were suggested to be spin-superfluid and spin-supersolid states, respectively \cite{liu1973quantum,fisher1974spin,tsurkan2017ultra}. Very recently, multiferroicity has been reported in the YK and intermediate phases of MnCr$_2$S$_4$ \cite{ruff2019multiferroic,Wang2020experimental}. However, a detailed understanding of the magnetic structure in the field-induced phases in MnCr$_2$S$_4$ is still missing. Magnetic spinels with a single magnetic ion could be studied in details by use of bulk magnetization and neutron scattering measurements \mbox{\cite{matsuda2007spin,miyata2011magnetic}}. MnCr$_2$S$_4$, however, possesses two magnetic ions featured by the site-specific magnetic properties at the $A$ and $B$ sites. This makes bulk-magnetization studies quite challenging to reveal the microscopic nature of the field-induced phases.

In this work, we investigate the microscopic nature by performing soft x-ray magnetic circular dichroism (XMCD) measurements of a MnCr$_2$S$_4$ single crystal in pulsed magnetic fields along with macroscopic thermodynamic experiments. The element-selective XMCD study, complemented by the extracted entropy changes, reveals a complex sublattice-moment rotation at the manganese and chromium sites in fields up to 40 T. Furthermore, the observed magnetocaloric effect (MCE) together with ultrasound and specific-heat measurements reveal a tetracritical point (TP) owing to spin-lattice coupling in the $H$-$T$ phase diagram, which is a rather surprising finding in an isotropic Heisenberg magnet.

We performed various experiments including soft XMCD at the Mn and Cr $L_3$ edges as well as magnetocaloric effect complemented by specific-heat and ultrasound experiments. Details of the experimental methods are described in Sec. II of the Supplemental Material \mbox{\cite{yamamoto2020SM}}. We note the very weak magnetic anisotropy due to half-filled $e_g$ and $t_{2g}$ (Mn$^{2+}$) and $t_{2g}$ (Cr$^{3+}$) states with zero orbital moment. This ensures that the present results in high magnetic fields are independent of the field direction. This was verified in previous studies that showed no magnetic anisotropy above 0.5 T \cite{tsurkan2002anomalous,tsurkan2003magnetic}.

\begin{figure}[t]
	\begin{center}
		\includegraphics[width=8.4cm]{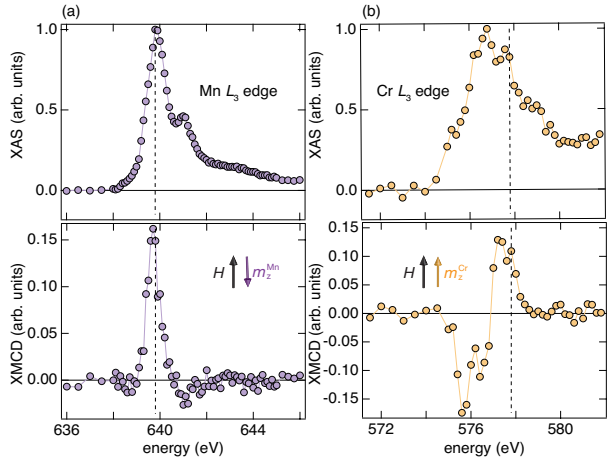}
	\end{center}
	\caption{XAS and XMCD spectra for the (a) Mn and (b) Cr $L_3$ edge. These spectra were obtained in zero field at 15 K after a field pulse of 20 T was applied along the [110] axis to induce a single-magnetic domain state. The vertical dashed lines represent the photon energies at which magnetic-field-dependent XMCD data were recorded [Figs. \ref{FigTwo}(b) and \ref{FigThree}]. The spectra are normalized to the $L_3$ XAS peak intensity. The illustration shows the relation of the longitudinal sublattice magnetization, $m_z$, with respect to the initial-field direction, $H$, (see text for details).}
	\label{FigOne}
\end{figure}


\begin{figure}[t]
	\begin{center}
		\includegraphics[width=6.8cm]{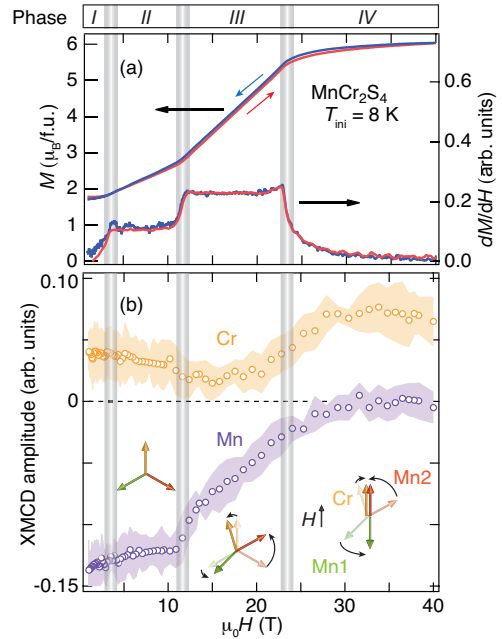}
	\end{center}
	\caption{(a) Field dependence of the bulk magnetization (left scale) and its field derivative (right scale). (b) Mn (purple) and Cr (orange) XMCD amplitudes are shown by symbols as function of magnetic field. The sign of the Mn XMCD is reversed for correspondence to the relative Mn sublattice magnetization. The shaded areas indicate the experimental errors. All data were collected at the initial temperature of 8 K. The vertical lines separate the phases $I$-$I\hspace{-.1em}V$, which are indicated on top of the figure. Schematic diagrams of the spin structures at the respective phase $I\hspace{-.1em}I$, $I\hspace{-.1em}I\hspace{-.1em}I$, and $I\hspace{-.1em}V$ are illustrated with orange for Cr, and green (red) arrows for Mn1 (Mn2). See text for details.}
	\label{FigTwo}
\end{figure}

\begin{figure}[t]
	\begin{center}
		\includegraphics[width=4.5cm]{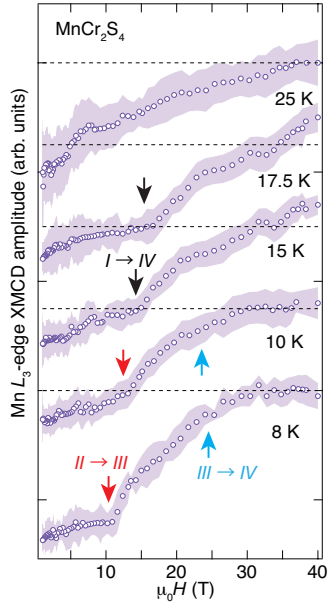}
	\end{center}
	\caption{Field dependence of the Mn $L_3$-edge XMCD amplitude with the experimental errors shown by the shaded area at selected initial temperatures $T_\mathrm{ini}$. The sign of the XMCD amplitudes is inverted to provide correspondence with the relative Mn sublattice magnetization. The data for $T_\mathrm{ini}$ = 8 K are the same as that shown in Fig. \ref{FigTwo}(b). The results for the field scans between 10 and 25 K are vertically shifted for clarity. The horizontal dashed lines show the zero level, which is close to the maximum field at each temperature. The red, blue, and black arrows denote the anomalies, which signal the phase transitions of $I\hspace{-.1em}I$-$I\hspace{-.1em}I\hspace{-.1em}I$, $I\hspace{-.1em}I\hspace{-.1em}I$-$I\hspace{-.1em}V$, and $I$-$I\hspace{-.1em}V$, respectively.}
	\label{FigThree}
\end{figure}

Figure \ref{FigOne} shows zero-field X-ray absorption spectroscopy (XAS) and XMCD results at the Mn and Cr $L_3$ edge measured at 15 K. In order to prepare a single-magnetic domain, a pulsed field of up to 20 T along the [110] axis (hereafter called initial field) was applied before taking the zero-field spectra. The XAS spectra with multiplet structures agree with those previously reported for Mn$^{2+}$ in tetrahedral \cite{hwang2007x} and Cr$^{3+}$ in octahedral symmetry \cite{kimura2001soft,deb2003soft}. Note that the Mn XAS and XMCD give an average information on the two nonequivalent Mn sublattices. The largest XMCD features near the $L_3$ edge (Mn at $\sim$ 639.8 eV and Cr at $\sim$ 576 eV) show an opposite sign of the XMCD amplitude, documenting the antiparallel coupling between the Mn and Cr 3$d$ longitudinal moments ($m_z$). In static-field bulk magnetization measurements, no remanence has been observed at temperatures above 5 K \cite{tsurkan2003magnetic}. The zero-field XMCD results imply that finite moments of the Mn and Cr sublattice are oriented antiparallel to each other in the initial-field axis, with the net moment is zero. The dashed line in Fig. \ref{FigOne}(a) [Fig. \ref{FigOne}(b)] indicates the photon energy of 639.8 eV (577.8 eV), which was chosen for studying the Mn (Cr) sublattice magnetization in pulsed-field XMCD experiments shown in Figs. \mbox{\ref{FigTwo}}(b) and \mbox{\ref{FigThree}}. The positive XMCD amplitude at these photon energies corresponds to the (anti)parallel Cr (Mn) 3$d$ magnetic moment $m_z^{\mathrm{Cr}}$ ($m_z^{\mathrm{Mn}}$) with respect to the field direction.

\begin{figure}[t]
	\begin{center}
		\includegraphics[width=5.2cm]{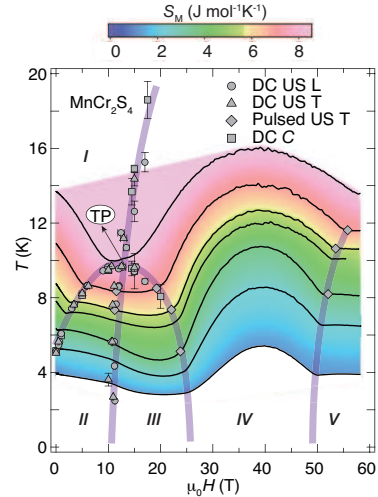}
	\end{center}
	\caption{Color-coded magnetic entropy $S_M$ in the $H$-$T$ plane, extracted from isentropic temperature variations in magnetic field shown by the black solid lines. The phase boundaries as determined by static (DC) and pulsed-field ultrasound (US) experiments for longitudinal (L) and transverse (T) acoustic modes, as well as from specific-heat ($C$) experiments are shown by symbols. The purple lines are guides to the eye and separate the phases $I$-$V$.}
	\label{FigFour}
\end{figure}

Figures \ref{FigTwo}(a) and \ref{FigTwo}(b) show bulk magnetization data and element-selective $L_3$-edge XMCD up to 40 T, respectively, recorded at the initial temperature $T_\mathrm{ini}$ of 8 K in pulsed magnetic fields. The XMCD amplitude is proportional to the total moment ($m_z$) of the respective atom \cite{van2014x,yamamoto2020element}. The sign of the Mn XMCD amplitude at 639.8 eV is reversed to provide correspondence with the direction of the moment of the Mn sublattice with respect to the field direction as shown in the illustration of Fig. \ref{FigOne} and which is also valid for Fig. \ref{FigThree}. The bulk magnetization and its field derivative signal a sequence of successive phase transitions ranging from $I$ to $I\hspace{-.1em}V$ as shown at the top of Fig. \ref{FigTwo}. These phases are suggested to be $I$ (ferrimagnetic order with disordered transverse components of the moments of the Mn ions), $I\hspace{-.1em}I$ (YK state), $I\hspace{-.1em}I\hspace{-.1em}I$ (low-field asymmetric state), and $I\hspace{-.1em}V$ (plateau state) \cite{tsurkan2017ultra}. 

In Ref. \cite{tsurkan2017ultra}, the low-field asymmetric state was described by a spin configuration with a tilted triangular structure, when compared to the YK configuration, however, with the chromium moments still aligned strictly parallel to the external magnetic field. In earlier work, this asymmetric spin state was thought to result from the manganese spins located at the two nonequivalent tetrahedral sites bearing magnetic moments of different length and sign \cite{denis1970magnetic} or this spin state was described by an ``oblique" spin configuration with deviations of the chromium moments from the external field direction, a configuration which was thought to be stabilized by biquadratic exchange \cite{plumier1980magnetic}. Similar spin structures, with the chromium moments not parallel to the external field were proposed and were shown to result from strong spin-lattice coupling effects \cite{miyata2020spin}. 

Figure \ref{FigTwo}(b) shows that the Cr and Mn XMCD amplitudes remain almost constant in phase $I$ and $I\hspace{-.1em}I$. The Mn XMCD amplitude markedly starts to increase above $\sim$ 11.5 T in phase $I\hspace{-.1em}I\hspace{-.1em}I$, which reflects the simultaneous rotation of the Mn1 and Mn2 moments and indicate a continuous increase of the opening angle between Mn1 and Mn2 moments. This finally leads to the collinear AFM state of the Mn1 and Mn2 spins along the field direction [110] in phase $I\hspace{-.1em}V$ where the Mn XMCD amplitude becomes zero.  On the other hand, the Cr XMCD amplitude decreases slightly in phase $I\hspace{-.1em}I\hspace{-.1em}I$ (Fig. \ref{FigTwo}). This could indeed reflect the deviation of the Cr magnetic moments from the original direction parallel to the field in the phase $I\hspace{-.1em}I$, which was predicted for the magneto-structural phase in a recent study and was explained by strong spin-lattice coupling \cite{miyata2020spin}. In phase $I\hspace{-.1em}V$, the Cr XMCD amplitude further increases and reaches its saturation significantly larger than the amplitude in phase $I\hspace{-.1em}I$. Monte-Carlo simulation and the bulk magnetization results  \mbox{\cite{miyata2020spin}} show that the size of the Cr moments are expected to be the same in phases $I$, $I\hspace{-.1em}I$, and $I\hspace{-.1em}V$. This suggests that the XMCD amplitude in phase $I\hspace{-.1em}V$ contains both contributions from local-structural and Cr magnetic-moment changes. The magnetic structures deduced from the XMCD data are shown in Fig. \ref{FigTwo}(b), which provides element-selective information on the proposed YK ($I\hspace{-.1em}I$), low-field asymmetric ($I\hspace{-.1em}I\hspace{-.1em}I$), and plateau ($I\hspace{-.1em}V$) states. As discussed in the following, this microscopic picture is further supported by our pulsed-field MCE measurements (Fig. \ref{FigFour}).


In the temperature range below 20 K, the moment of the Mn sublattice is anticipated to be responsible for the main contributions to the field-induced changes in the bulk magnetization, rather than the Cr counterpart \cite{tsurkan2003magnetic}. Figure \ref{FigThree} shows the field dependence of the Mn $L_3$-edge XMCD amplitude at selected $T_\mathrm{ini}$. At $T_\mathrm{ini}$ = 8 and 10 K, the system evolves through the phases $I$, $I\hspace{-.1em}I$, $I\hspace{-.1em}I\hspace{-.1em}I$, and $I\hspace{-.1em}V$ as shown in Fig. \ref{FigFour}. The field-dependent XMCD amplitude signals the transitions of $I\hspace{-.1em}I$-$I\hspace{-.1em}I\hspace{-.1em}I$ and $I\hspace{-.1em}I\hspace{-.1em}I$-$I\hspace{-.1em}V$. At $T_\mathrm{ini}$ = 15 and 17.5 K, anomalies are identified that correspond to the phase transition from $I$ to $I\hspace{-.1em}V$. At $T_\mathrm{ini}$ = 25 K, no anomaly is detected up to 40 T, which well agrees with the previous magnetization data \cite{tsurkan2017ultra}. This confirms that the observed variations of the Mn XMCD signals originate from the Mn 3$d$ magnetic moment. We note that Monte Carlo simulation suggests subtle displacements of the sulfur atoms that bridge the Mn and Cr cations \mbox{\cite{miyata2020spin}}. In future, it will be important to investigate the lineshapes of XAS/XMCD spectra at high magnetic fields in phase $I\hspace{-.1em}I\hspace{-.1em}I$ and $I\hspace{-.1em}V$, which could provide further insights into such subtle changes of the local environments.


Now, we present the thermodynamic results. Figure \ref{FigFour} shows the contour plot of the magnetic entropy $S_\mathrm{M}$ extracted from the isentropic $T(H)$ data (black solid lines) obtained in pulsed fields up to 60 T. The phase boundaries are drawn by purple lines deduced from ultrasound and specific-heat anomalies (see Sec. III of the Supplemental Material \mbox{\cite{yamamoto2020SM}}), which separate the phases $I$-$V$ (see Sec. I of the Supplemental Material \mbox{\cite{yamamoto2020SM}}). The absolute values of the entropy were calibrated with data at 10 T from literature \cite{tsurkan2003magnetic}. 

Each MCE trace shows multiple kinks that signal successive magnetic phase transitions. Obviously, the MCE exhibits the anomalies at the same positions as those deduced from ultrasound and specific-heat measurements. Starting from the phase $I$, the temperature decreases in the adiabatic processes due to the increase of $S_M$ toward the phases $I\hspace{-.1em}I$ and $I\hspace{-.1em}I\hspace{-.1em}I$. After entering the phase $I\hspace{-.1em}V$, the adiabatic temperature remarkably increases resulting from the decrease of $S_M$ up to 40 T. Then $S_M$ increases toward the phase $V$ and levels off. The overall  phase boundaries are qualitatively in good agreement with the previous works based on other macroscopic experiments \cite{tsurkan2017ultra,miyata2020spin}. The smaller entropy values $S_M$ in the phase $I\hspace{-.1em}V$ compared to the phases $I\hspace{-.1em}I$ and $I\hspace{-.1em}I\hspace{-.1em}I$, probably signal the larger noncollinearity of the magnetic structures in the latter states. 

Our element-selective XMCD results complemented by this MCE data provide compelling evidence for the proposed noncollinear (collinear) magnetic structures in the phases $I\hspace{-.1em}I$ and $I\hspace{-.1em}I\hspace{-.1em}I$ ($I\hspace{-.1em}V$). This is in line with a recent report suggesting a spin current or inverse Dzyaloshinskii-Moriya mechanism to be responsible for the multiferroic properties observed in the phases $I\hspace{-.1em}I$ and $I\hspace{-.1em}I\hspace{-.1em}I$ \mbox{\cite{ruff2019multiferroic}}. Furthermore, our MCE results indicate the importance of the adiabatic temperature changes for mapping the $H$-$T$ phase diagram from pulsed-field experiments (see Sec. IV of the Supplemental Material \mbox{\cite{yamamoto2020SM}}).

Remarkably, our observations signal the presence of a tetracritical point in the phase diagram, which was already suggested from the phase diagram determined from pyrocurrent measurements \cite{ruff2019multiferroic}. The TP has been theoretically predicted \cite{liu1973quantum,fisher1974spin,bruce1975coupled,kosterlitz1976bicritical} for AFM systems with competing magnetic anisotropies, and experimentally observed as well \cite{bevaart1979neutron,shapira1981tetracriticality}. On the other hand, MnCr$_2$S$_4$ has three spin degrees of freedom ($n$ = 3), dominated by rather isotropic Mn-Mn, Mn-Cr, and Cr-Cr Heisenberg exchange interactions. According to renormalization-group theory, the TP does not emerge in $n$ = 3 spin systems under uniform or staggered magnetic fields with either longitudinal or transverse component \mbox{\cite{fisher1974spin,kosterlitz1976bicritical}}. However, very recently, a TP was analytically shown to be one of the possible three nondecomposable critical points for systems with $n$ = 3 using renormalization group theory in $d$ = 4-$\varepsilon$ dimensions \mbox{\cite{codello2020critical}}. In MnCr$_2$S$_4$, the intermediate phase ($I\hspace{-.1em}I\hspace{-.1em}I$) is stabilized via spin-lattice coupling (see Sec. V of the Supplemental Material \mbox{\cite{yamamoto2020SM}}). Monte-Carlo calculations show that strong spin-lattice interaction in MnCr$_2$S$_4$ spontaneously breaks the $Z_2$ symmetry of the Mn-Cr bonds in the phases $I\hspace{-.1em}I\hspace{-.1em}I$, $I\hspace{-.1em}V$, and $V$ \cite{miyata2020spin} and effectively generates staggered magnetic fields at the site of the manganese spins. The effective staggered field in MnCr$_2$S$_4$ has both longitudinal and transverse components with respect to the uniform external magnetic field, originating from a slight canting of the Cr moment observed in our experiments [Fig. \mbox{\ref{FigTwo}}(b)]. This stabilizes the asymmetric intermediate phase and results in the TP, which is rather unusual for conventional Heisenberg systems \cite{kosterlitz1976bicritical}.

Finally, we discuss the symmetry of the phases $I$--$I\hspace{-.1em}V$ based on the magnetic structures deduced from the current study. At zero field, with decreasing temperature, the O(3) and O(2) symmetries are successively broken at $T_C$ and $T_\mathrm{YK}$, respectively. In finite magnetic fields, the O(3) symmetry is always broken. In the low-field intermediate phase $I\hspace{-.1em}I\hspace{-.1em}I$, both the O(2) and $Z_2$ symmetries are broken. On the other hand, in the plateau phase $I\hspace{-.1em}V$, the $Z_2$ symmetry is kept broken, while the O(2) symmetry is restored. A series of symmetry breakings at the respective phase transitions is consistent with the earlier theoretical studies which predict the TP \mbox{\cite{fisher1974spin,kosterlitz1976bicritical}}.

In summary, element-specific XMCD together with the observed entropy variations provide direct evidence of the proposed noncollinear magnetic structures in the YK ($I\hspace{-.1em}I$), low-field asymmetric ($I\hspace{-.1em}I\hspace{-.1em}I$), and plateau ($I\hspace{-.1em}V$) phases. Because the $H$-$T$ phase diagram of MnCr$_2$S$_4$ is symmetric with respect to a field of approximately 40 T \cite{tsurkan2017ultra,miyata2020spin}, the present results provide a plausible description of the magnetic structures beyond 40 T, in phase $V$, and the higher-field phase (called inverse YK phase in Ref. \cite{miyata2020spin}) close to saturation. In addition, our thermodynamic results hint toward an alternative route to the tetracritical point (TP) due to a novel mechanism, significant spin-lattice couplings, in frustrated systems. We believe the understanding of the relation between the TP and spin-strain interactions might give further insight into the spin-superfluid and supersolid states suggested in analogy to bosonic systems \cite{matsuda1970off,tsurkan2017ultra}.

We acknowledge the support of the HLD at HZDR, member of the European Magnetic Field Laboratory (EMFL), the Deutsche Forschungsgemeinschaft (DFG) through SFB 1143, the W\"{u}rzburg-Dresden Cluster of Excellence on Complexity and Topology in Quantum Matter--$ct.qmat$ (EXC 2147, Project ID 390858490), and the BMBF via DAAD (Project ID 57457940). This work was partly supported by the DFG through Transregional Research Collaboration TRR 80 (Augsburg, Munich, and Stuttgart) as well as by the project ANCD 20.80009.5007.19 (Moldova). This work was partly performed at High Field Laboratory for Superconducting Materials, Institute for Materials Research, Tohoku University (Project No 19H0505). The synchrotron radiation experiments were performed at the BL25SU of SPring-8 with the approval of the Japan Synchrotron Radiation Research Institute (JASRI) (Proposal No. 2019A1534, 2019B1474). We would like to thank A. Miyata for fruitful discussions and P. T. Cong for technical support.

\end{document}


\enlargethispage{2\baselineskip}
\title{Supplemental Material\\Element-specific field-induced spin reorientation and \\an unusual tetracritical point in MnCr$_2$S$_4$}

\author{Sh. Yamamoto}
\email{s.yamamoto@hzdr.de}
\affiliation{Hochfeld-Magnetlabor Dresden (HLD-EMFL) and W\"{u}rzburg-Dresden Cluster of Excellence ct.qmat, Helmholtz-Zentrum Dresden-Rossendorf, 01328 Dresden, Germany}
\author{H. Suwa}
\affiliation{Department of Physics, The University of Tokyo, Tokyo 113-0033, Japan}
\author{T. Kihara}
\affiliation{Institute for Materials Research, Tohoku University, Sendai 980-8577, Japan}
\author{T. Nomura}
\affiliation{Hochfeld-Magnetlabor Dresden (HLD-EMFL) and W\"{u}rzburg-Dresden Cluster of Excellence ct.qmat, Helmholtz-Zentrum Dresden-Rossendorf, 01328 Dresden, Germany}
\affiliation{Institute for Solid State Physics, The University of Tokyo, Kashiwa, Chiba 277-8581, Japan}
\author{Y. Kotani}
\affiliation{Japan Synchrotron Radiation Research Institute, SPring-8, Sayo, Hyogo 679-5198, Japan}
\author{T. Nakamura}
\affiliation{Institute for Materials Research, Tohoku University, Sendai 980-8577, Japan}
\affiliation{Japan Synchrotron Radiation Research Institute, SPring-8, Sayo, Hyogo 679-5198, Japan}
\author{Y. Skourski}
\affiliation{Hochfeld-Magnetlabor Dresden (HLD-EMFL) and W\"{u}rzburg-Dresden Cluster of Excellence ct.qmat, Helmholtz-Zentrum Dresden-Rossendorf, 01328 Dresden, Germany}
\author{S. Zherlitsyn}
\affiliation{Hochfeld-Magnetlabor Dresden (HLD-EMFL) and W\"{u}rzburg-Dresden Cluster of Excellence ct.qmat, Helmholtz-Zentrum Dresden-Rossendorf, 01328 Dresden, Germany}
\author{L. Prodan}
\affiliation{Institute of Applied Physics, MD 2028, Chisinau, R. Moldova}
\affiliation{Experimental Physics 5, Center for Electronic Correlations and Magnetism,
Institute of Physics, University of Augsburg, 86159, Augsburg, Germany}
\author{V. Tsurkan}
\affiliation{Institute of Applied Physics, MD 2028, Chisinau, R. Moldova}
\affiliation{Experimental Physics 5, Center for Electronic Correlations and Magnetism,
Institute of Physics, University of Augsburg, 86159, Augsburg, Germany}
\author{H. Nojiri}
\affiliation{Institute for Materials Research, Tohoku University, Sendai 980-8577, Japan}
\author{A. Loidl}
\affiliation{Experimental Physics 5, Center for Electronic Correlations and Magnetism,
Institute of Physics, University of Augsburg, 86159, Augsburg, Germany}
\author{J. Wosnitza}
\affiliation{Hochfeld-Magnetlabor Dresden (HLD-EMFL) and W\"{u}rzburg-Dresden Cluster of Excellence ct.qmat, Helmholtz-Zentrum Dresden-Rossendorf, 01328 Dresden, Germany}
\affiliation{Institut f$\ddot{u}$r Festk$\ddot{o}$rper- und Materialphysik, TU Dresden, 01062 Dresden, Germany}

\date{\today}
\maketitle

\section{Predicted magnetic structures in magnetic-field-induced phases}

\begin{figure}[H]
	\begin{center}
		\includegraphics[width=17cm]{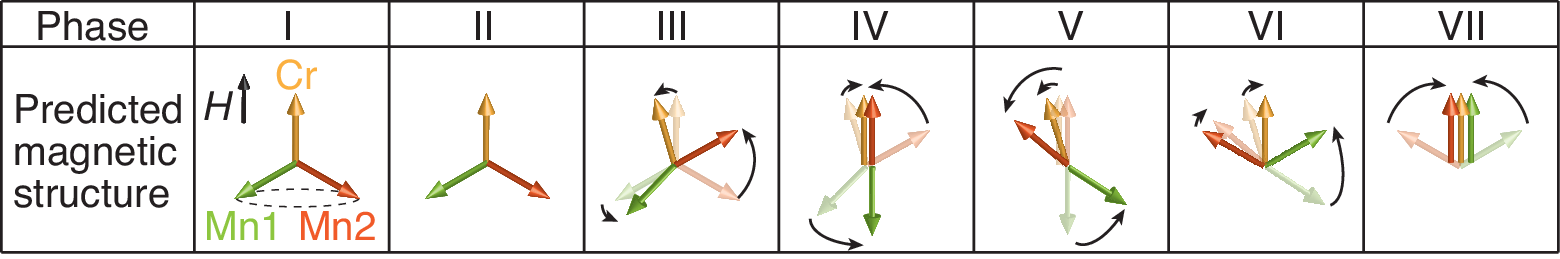}
	\end{center}
	\caption{Schematic diagram of the magnetic structures predicted in earlier studies \cite{tsurkan2017ultra,miyata2020spin}. Phase $I$: ferrimagnetic order with disordered transverse components of the moments of the Mn ions, $I\hspace{-.1em}I$: YK state, $I\hspace{-.1em}I\hspace{-.1em}I$: low-field asymmetric state, $I\hspace{-.1em}V$: plateau state, $V$: high-field asymmetric state, $V\hspace{-.1em}I$: inverse YK state, and $V\hspace{-.1em}I\hspace{-.1em}I$: forced ferromagnetic state. The notation of the phases $I$ -- $V$ corresponds to those in Fig. 4. The phases $V\hspace{-.1em}I$ and $V\hspace{-.1em}I\hspace{-.1em}I$ were suggested in Ref. \cite{miyata2020spin}.}
	\label{FigS1}
\end{figure}

Fig. \mbox{\ref{FigS1}} shows predicted magnetic structures for MnCr$_2$S$_4$ composed of the chromium moment at the $A$ site and two nonequivalent manganese moments at the $B$ site in the field-induced phases reported in earlier studies \cite{tsurkan2017ultra,miyata2020spin}. The phases of $I$ -- $V$ are located up to 60 T below 20 K (Fig. 4 in the main text). The phases $V\hspace{-.1em}I$ and $V\hspace{-.1em}I\hspace{-.1em}I$ are reached at a magnetic field of 75 and 85 T, respectively \cite{miyata2020spin}.

\section{Experimental methods}

X-ray absorption spectroscopy (XAS) and XMCD experiments were performed at the soft x-ray twin-helical undulator beamline BL25SU \cite{senba2016upgrade} in SPring-8 under pulsed fields up to 40 T applied along the [110] axis with an energy resolution of $E/\Delta E=3000$. A solenoid magnet with a bore size of 18 mm was used with a capacitor bank with a charging voltage of 3.1 kV, and a capacitance of 16 mF. The pulse duration with the magnetic field larger than 10 \% of the maximum field, was 22 ms. The total-electron yield (TEY) of $\mu_{\mathrm{+}}$ ($\mu_{\mathrm{-}}$) under pulsed fields was measured for right (left) circularly polarized light at the Mn and Cr $L_3$ absorption edge (2$p_{3/2}\ \to$ 3$d$) by using a time-resolved acquisition technique \cite{nakamura2011soft,yamamoto2020element}. The XAS and XMCD signals were obtained from ($\mu_{\mathrm{+}}$+$\mu_{\mathrm{-}}$)/2, and $\mu_{\mathrm{+}}$--$\mu_{\mathrm{-}}$, respectively. The sample was cleaved $in$ $situ$ at pressures below 2 $\times$ 10$^{-7}$ Pa. The field dependence of the Mn and Cr $L_3$-edge XMCD amplitude [Figs. 2(b) and 3 in the main text] was obtained by averaging results of 14 field pulses. The error bars were extracted from standard deviations.

Ultrasound measurements were conducted using a pulse-echo phase-sensitive detection technique \cite{zherlitsyn2014spin} in static fields up to 17 T, and in pulsed fields up to 60 T with the sound propagation vector, $\bm{\mathrm{k}}$, parallel to [111]. Both, longitudinal ($\bm{\mathrm{k}}\ ||\ \bm{\mathrm{u}}$) and transverse ($\bm{\mathrm{k}}\hspace{-.4em}\ \perp\ \hspace{-.4em}\bm{\mathrm{u}}$) modes with $\bm{\mathrm{u}}||$[1$\overline{1}$0] were studied, where $\bm{\mathrm{u}}$ is the displacement vector. These modes are related to the  elastic constants $c_L = (c_{11}+2c_{12}+4c_{44})/3$ and $c_T = (c_{11}+c_{44}-c_{12})/3$, respectively. In the static-field measurements, the ultrasound frequency below (above) 12 T was 60.7 (55.6) MHz. Specific-heat ($C$) measurements were performed using a 25 T cryogen-free superconducting magnet. $C$ vs $T$ was obtained using both, a standard relaxation technique in fields above 17 T and a dual-slope method \cite{riegel1986dual} in fields below 14.5 T. MCE in pulsed magnetic fields was measured by using a RuO$_2$ thermometer (900 $\si{\ohm}$, 0.6 $\times$ 0.3 $\times$ 0.1 mm$^3$) under adiabatic conditions \cite{kihara2013adiabatic,nomura2017alpha}. A standard ac four-probe method with a numerical lock-in technique at a frequency of 50 kHz was applied to measure the resistance of the thermometer. High-field magnetization was investigated using a coaxial pickup coil system \cite{skourski2011high}. Specific heat, MCE, ultrasound, and bulk magnetization measurements were performed in magnetic fields applied along the [111] direction.

\newpage
\section{Ultrasound and specific-heat results}

\begin{figure}[H]
	\begin{center}
		\includegraphics[width=10.5cm]{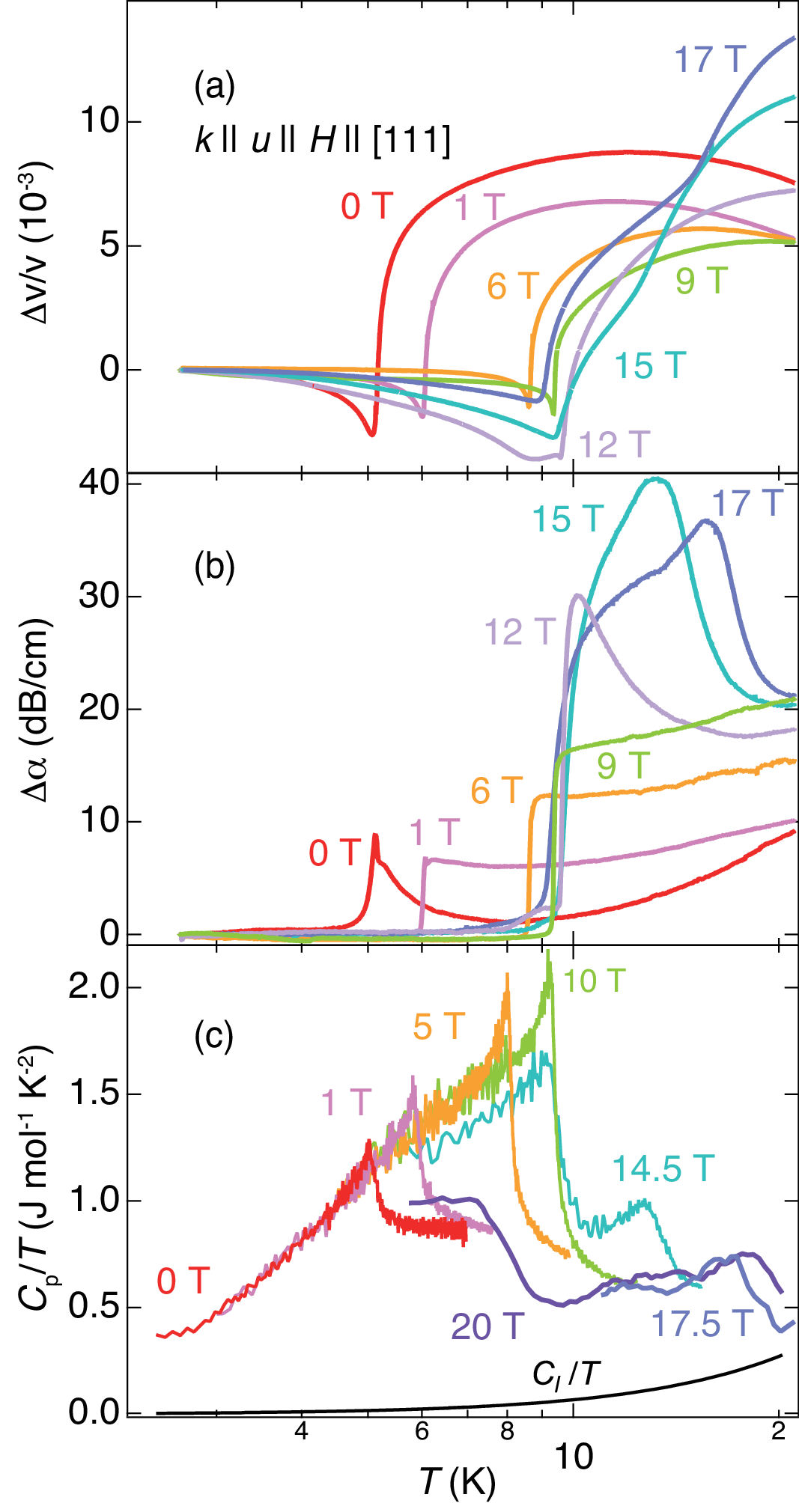}
	\end{center}
	\caption{(a) Relative sound-velocity changes $\Delta v/v$ and (b) the sound attenuation $\Delta \alpha$ of the longitudinal acoustic mode propagating along the [111] direction as a function of temperature, plotted on a  semilogarithmic scale for selected magnetic fields. (c) Temperature dependence of the molar specific heat $C$ divided by temperature at various magnetic fields. The estimated lattice contribution $C_l$/$T$ is shown by a black solid line. }
	\label{FigS2}
\end{figure}

In Figs. \ref{FigS2}(a) and \ref{FigS2}(b), the relative sound velocity $\Delta v/v$ and sound attenuation $\Delta \alpha$ are plotted on a logarithmic temperature scale for the longitudinal acoustic mode at different static magnetic fields. The strong exchange-striction coupling makes the ultrasonic properties sensitive to the maganese spin structures \cite{tsurkan2017ultra,miyata2020spin}. At zero field, a well-defined deep minimum in $\Delta v/v$ and a $\lambda$-type anomaly in $\Delta \alpha$ appear at $T$ $\approx$ 5 K. These anomalies shift to higher temperatures with increasing field, which matches well with the field dependence of $T_\mathrm{{YK}}$ previously reported \cite{tsurkan2003magnetic, tsurkan2017ultra}. Above 12 T, where phase $I\hspace{-.1em}I\hspace{-.1em}I$ is the zero-temperature ground state, the anomalies are seen near $T$ $\approx$ 9.5 K, representing the high-field branch of the phase boundary recently suggested \cite{ruff2019multiferroic}. Above 15 T, $\Delta v/v$ ($\Delta \alpha$) shows an additional kink (broad maximum) at higher temperatures between 12 and 16 K.

In Fig. \ref{FigS2}(c), the temperature dependence of the molar specific heat $C$ is shown in fields up to 20 T, together with the lattice contribution, $C_l$, by assuming a Debye temperature of 410 K \cite{tsurkan2003magnetic}. Earlier conductivity studies reported negligible electronic contributions to $C$ in this temperature range \cite{tsurkan2003magnetic}. Therefore, magnetic degrees of freedom play the dominant role in the variation of $C$. The $\lambda$-type anomaly is shifted to higher temperatures with increasing fields up to 10 T. At fields beyond 14.5 T, additional anomalies are seen in the temperature range from 8 to 19 K. The specific-heat results essentially capture the same phase boundaries compared to those observed in the ultrasonic experiments.

\begin{figure}[H]
	\begin{center}
		\includegraphics[width=12.5cm]{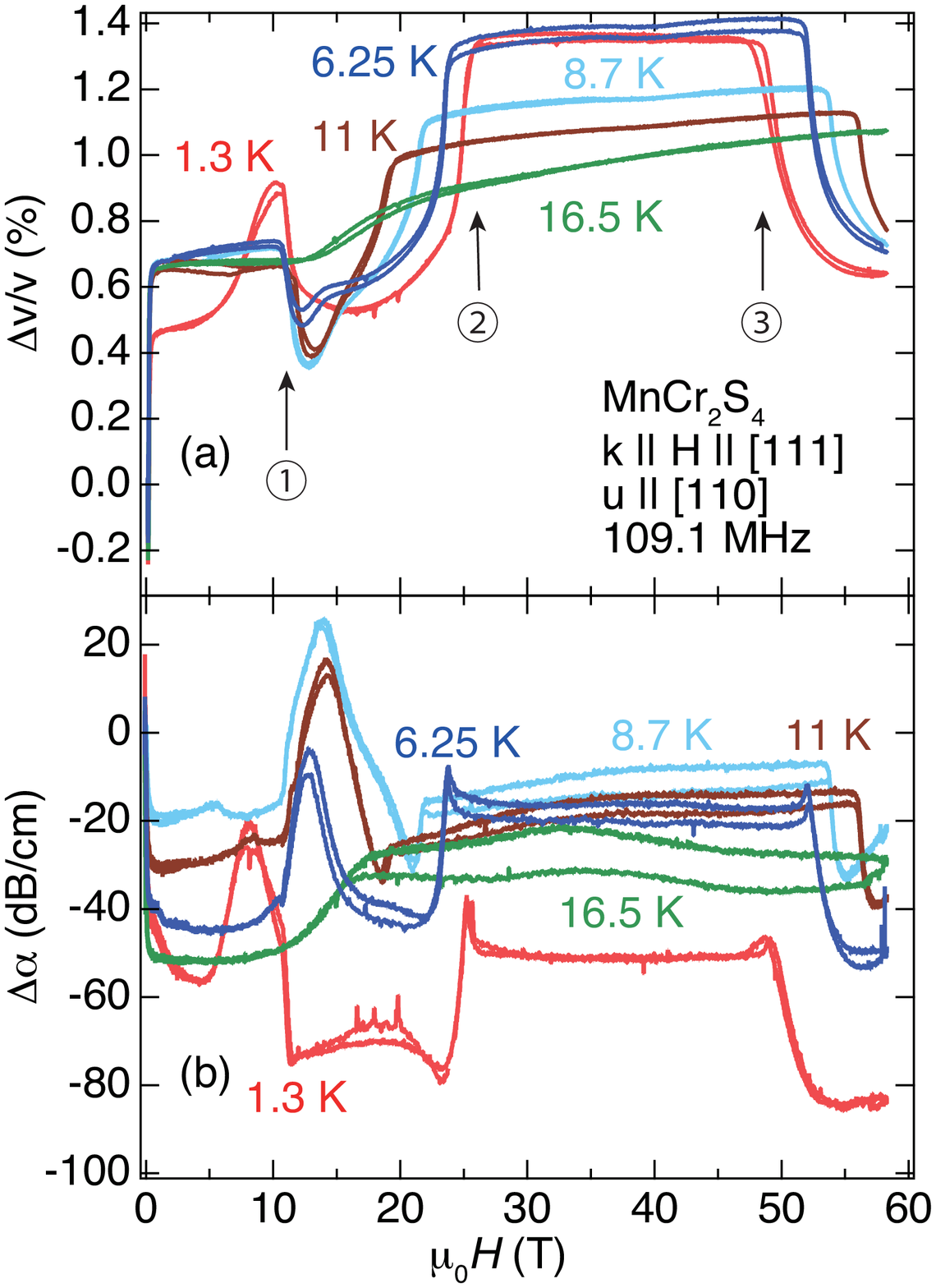}
	\end{center}
	\caption{(a) The relative sound-velocity change, $\Delta v/v$, and (b) sound attenuation, $\Delta \alpha$, of the transverse wave propagating along the [111] direction ($\bm{k}||\bm{H}||$[111], $\bm{u}||$[1$\overline{1}$0]) as function of magnetic field for selected initial temperatures. The ultrasound frequency was 109.1 MHz. Arrows indicate field-induced phase transitions of 1: $I\hspace{-.1em}I$--$I\hspace{-.1em}I\hspace{-.1em}I$, 2: $I\hspace{-.1em}I\hspace{-.1em}I$--$I\hspace{-.1em}V$, and 3: $I\hspace{-.1em}V$--$V$, respectively.}
	\label{FigS3}
\end{figure}

Figure \ref{FigS3} shows the acoustic properties of the transverse mode ($\bm{k}||\bm{H}||$[111], $\bm{\mathrm{u}}||$[1$\overline{1}$0]) in pulsed magnetic fields up to 60 T for selected initial temperatures $T_\mathrm{ini}$. At $T_\mathrm{ini}$ = 1.3 K, the relative sound-velocity change, $\Delta v/v$, exhibits a drop by 0.3 \% at $\sim$ 11 T (anomaly 1), followed by a further decrease. After reaching the minimum at 16 T, it shows a hardening of this transverse mode by 0.8 \% with the maximum slope near 25 T (anomaly 2). $\Delta v/v$ shows a plateau up to 48 T, which ends with a step-like decrease (anomaly 3). Above 48 T, there is a gradual softening up to the highest fields. The anomalies 1, 2, and 3 signal the field-induced phase transitions of $I\hspace{-.1em}I$--$I\hspace{-.1em}I\hspace{-.1em}I$, $I\hspace{-.1em}I\hspace{-.1em}I$--$I\hspace{-.1em}V$, and $I\hspace{-.1em}V$--$V$, respectively. With increasing temperature, the anomaly 1 stays at almost the same location, whereas the anomaly 2 (3) shifts to the lower (higher) magnetic fields. These anomalies are located at the same position as those obtained from the longitudinal-mode results \cite{tsurkan2017ultra}. The changes in $\Delta v/v$ at anomalies are about one order of magnitude smaller than those observed for the longitudinal mode. This indicates larger magnetoelastic coupling for the latter mode. The sound attenuation $\Delta \alpha$ correspondingly shows a series of anomalies as well. The anomalies 1 -- 3, which appear in both, $\Delta v/v$ and $\Delta \alpha$ simultaneously were plotted in the $H$-$T$ phase diagram (Fig. 4 in the main text). The longitudinal mode exhibits a softening centered at 40 T and a suppression of the sound attenuation (called anomaly 4 in Ref. \cite{tsurkan2017ultra}) above 6 K, which were not reproduced in classical Monte Carlo calculations \cite{miyata2020spin}. In the transverse mode, no anomalies near 40 T have been observed [Fig. \ref{FigS3}]. Further theoretical studies which include corresponding spin-strain interactions may help to understand the magnetoacoustic properties in the plateau region $I\hspace{-.1em}I\hspace{-.1em}I$. 

\newpage
\section{MCE and the $H$-$T$ phase diagram}

The $H$-$T$ phase diagram constructed from the pulsed-field results obtained under (quasi)adiabatic conditions can lead to shifted phase boundaries if the adiabatic temperature changes caused by magnetocaloric effect (MCE) are not properly taken into account \cite{brambleby2017adiabatic}. Earlier pulsed-field measurements indicated that the phase boundaries of $I$-$I\hspace{-.1em}I$ and $I\hspace{-.1em}I\hspace{-.1em}I$-$I\hspace{-.1em}V$ extend up to $\sim$ 13 -- 14 K \cite{tsurkan2017ultra,ruff2019multiferroic,miyata2020spin}. However, the present MCE results provide corrected phase boundaries at lower temperatures, which forms the dome-like structure from the phases $I\hspace{-.1em}I$ and $I\hspace{-.1em}I\hspace{-.1em}I$ in the $H$-$T$ phase diagram [Fig. 4 in the main text]. These results imply that the recently suggested ferroelectric phase (called FE3 in Ref. \cite{ruff2019multiferroic}) resides in the plateau state, $I\hspace{-.1em}V$.

\begin{figure}[H]
	\begin{center}
		\includegraphics[width=15cm]{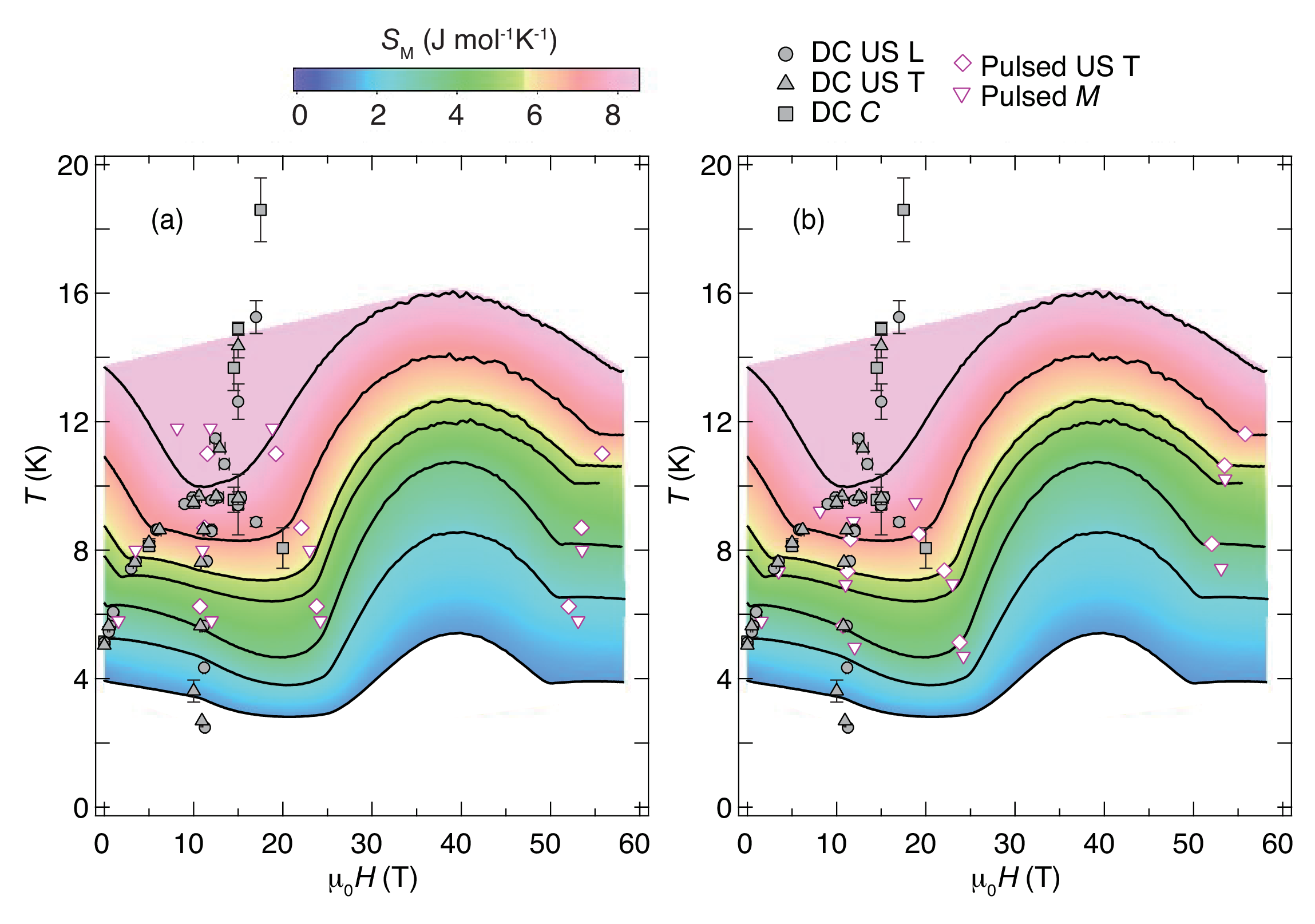}
	\end{center}
	\caption{$H$-$T$ phase diagram of MnCr$_2$S$_4$ with a color-coded map of magnetic entropy $S_M$. The closed symbols are obtained from static-field (DC) ultrasound (US) for the longitudinal (L) and transverse (T) modes, and specific-heat ($C$) measurements. The open symbols are taken from pulsed-field US measurements for the T mode and magnetization ($M$) measurements (a) without and (b) with taking the adiabatic temperature changes into account. } 
	\label{FigS4}
\end{figure}

Figure \ref{FigS4} shows the influence of the MCE effect on the phase diagram of MnCr$_2$S$_4$. The closed symbols are obtaind from static-field (DC) ultrasound (US) and specific-heat ($C$) measurements, while the open symbols are taken from pulsed-field US and magnetization ($M$) measurements. In Fig. \ref{FigS4}(a), initial temperatures $T_\mathrm{ini}$ are used for plotting the anomalies from the pulsed-field data (open symbols). On the other hand, in Fig. \ref{FigS4}(b), the temperature of each anomaly from the pulsed-field data was linearly interpolated using two of the seven MCE curves, which are shown by black solid lines. For example, for the pulsed-field magnetization data at $T_\mathrm{ini}$ = 11.8 K, we used the MCE results with $T_\mathrm{ini}$ = 10.9 K and $T_\mathrm{ini}$ = 13.7 K for linear interpolation. Each temperature from the magnetization data at a certain critical field ($H_c$) was linearly corrected using the MCE results. When the MCE is not considered [Fig. \ref{FigS4}(a)], additional phases seem to exist at higher temperatures than the phases $I\hspace{-.1em}I$ and $I\hspace{-.1em}I\hspace{-.1em}I$, which was assumed in Ref. \cite{ruff2019multiferroic}. Figure \ref{FigS4}(b) indicates that the anomalies obtained from the pulsed-field data map well onto the dome-like boundary that separates the phases $I$--$I\hspace{-.1em}I$ and $I\hspace{-.1em}I\hspace{-.1em}I$--$I\hspace{-.1em}V$ as measured under static-field condition. Furthermore, the anomalies at higher fields above 50 T are also located at positions closer to the kinks observed in the $T(H)$ curves in Fig. \ref{FigS4}(b) than those in Fig. \ref{FigS4}(a).

\newpage
\section{Spin-lattice coupling model}
We consider an effective model for MnCr$_{2}$S$_{4}$, in which Mn
and Cr ions have $S=5/2$ and $3/2$ spins, respectively. The Mn ions
form a bipartite diamond lattice consisting of the two sublattices
$\mathcal{A}$ and $\mathcal{B}$, and the Cr ions form a pyrochlore
lattice. This ferrimagnetic spinel exhibits successive phase transitions
under magnetic field. We found spin-lattice coupling essential to
this compound and proposed an effective model including bond phonons.
The Hamiltonian is represented by $\mathcal{H=H_{{\rm MM}}+H_{{\rm CC}}+H_{{\rm MC}}+H_{Z}}$
with
\begin{align}
\mathcal{H}_{{\rm MM}} & =\sum_{\langle ij\rangle}J_{{\rm Mn}\mbox{-}{\rm Mn}}{\bf S}_{{\rm Mn}i}\cdot{\bf S}_{{\rm Mn}j},\\
\mathcal{H}_{{\rm CC}} & =\sum_{\langle ij\rangle}J_{{\rm Cr}\mbox{-}{\rm Cr}}{\bf S}_{{\rm Cr}i}\cdot{\bf S}_{{\rm Cr}j},\\
\mathcal{H}_{{\rm MC}} & =\sum_{\langle ij\rangle}J_{{\rm Mn}\mbox{-}{\rm Cr}}\left(1-\alpha\rho_{ij}\right){\bf S}_{{\rm Mn}i}\cdot{\bf S}_{{\rm Cr}j}+\frac{K}{2}\rho_{ij}^{2},\\
\mathcal{H}_{{\rm Z}} & =-g\mu_{{\rm B}}{\bf B}\cdot\left(\sum_{i}{\bf S}_{{\rm Mn}i}+\sum_{j}{\bf S}_{{\rm Cr}j}\right).
\end{align}
If the spin-lattice coupling is absent $(\alpha=0)$, we can rewrite
the Hamiltonian $H=H_{{\rm tri}}+H_{{\rm tetra}}+{\rm const.}$ with
\begin{align}
H_{{\rm tri}} & =\sum_{{\rm \triangle}}\frac{J_{{\rm Mn}\mbox{-}{\rm Mn}}}{12}\left({\bf S}_{{\rm Mn\mathcal{A}}}+{\bf S}_{{\rm Mn}\mathcal{B}}+r{\bf S}_{{\rm Cr}}-\gamma{\bf B}\right)^{2},\\
H_{{\rm tetra}} & =\sum_{{\rm tetra}}\frac{J_{{\rm Cr}\mbox{-}{\rm Cr}}}{2}\left(\sum_{i\in{\rm tetra}}{\bf S}_{{\rm Cr}i}+\eta{\bf B}\right)^{2}.
\end{align}
Here the sums $\sum_{\triangle}$, $\sum_{{\rm tetra}}$, and $\sum_{i\in{\rm tetra}}$
are taken over all the triangles formed by connected three sites (Cr
site and two Mn sites in $\mathcal{A}$ and $\mathcal{B}$), over
all the tetrahedra consisting of the pyrochlore lattice, and over
all the four Cr sites for each tetrahedron, respectively. We define
$r=\frac{3J_{{\rm Mn}\mbox{-}{\rm Cr}}}{J_{{\rm Mn}\mbox{-}{\rm Mn}}}$,
$\gamma=\frac{g\mu_{{\rm B}}}{4J_{{\rm Mn}\mbox{-}{\rm Mn}}}$, and
$\eta=-\frac{g\mu_{{\rm B}}}{2J_{{\rm Cr}\mbox{-}{\rm Cr}}}\left(1+\frac{3J_{{\rm Mn}\mbox{-}{\rm Cr}}}{4J_{{\rm Mn}\mbox{-}{\rm Mn}}}\right)$.
The signs of the exchange couplings are $J_{{\rm Mn}\mbox{-}{\rm Mn}},J_{{\rm Mn}\mbox{-}{\rm Cr}}>0$,
$J_{{\rm Cr}\mbox{-}{\rm Cr}}<0$, and thus $r,\gamma,\eta>0$. We
can easily minimize the total energy by setting ${\bf S}_{{\rm Cr}i}\parallel{\bf B}$
for all the Cr spins and satisfying the following condition for each
triangle
\begin{equation}
{\bf S}_{{\rm Mn\mathcal{A}}}+{\bf S}_{{\rm Mn}\mathcal{B}}+r{\bf S}_{{\rm Cr}}-\gamma{\bf B}=0.
\end{equation}
As a result, the spin structure of the ground state is coplanar, and
$\left({\bf S}_{{\rm Mn\mathcal{A}}}+{\bf S}_{{\rm Mn}\mathcal{B}}\right)\parallel{\bf B}$.
The magnetization parallel to the magnetic field, $\frac{1}{2}\left({\bf S}_{{\rm Mn\mathcal{A}}}+{\bf S}_{{\rm Mn}\mathcal{B}}\right)+2{\bf S}_{{\rm Cr}}$,
is a monotonically increasing function of $|{\bf B}|$; no phase transition
occurs at zero temperature. Hence, the spin-lattice coupling is essential
to the experimentally observed plateau (and also the intermediate)
phase at low temperatures.